\documentstyle[epsf,twoside,fleqn,espcrc2]{article}

\newcommand{\AmS}{{\protect\the\textfont2
  A\kern-.1667em\lower.5ex\hbox{M}\kern-.125emS}}

\title{
\vspace{-4.8cm}
\begin{flushright}
{\normalsize KOMA-96-37}\\
{\normalsize August 1996}\\
\end{flushright}
\vspace*{3.0cm}
Softening Transitions with Quenched 2D Gravity}

\author{C.F. Baillie\address{Department of Computer Science,
         University of Colorado, Boulder, CO 80309, USA},
        W. Janke\address{Institut f\"ur Physik,
                    Johannes Gutenberg-Universit\"at Mainz, Staudinger Weg 7, 55099 Mainz, Germany }
       and 
        D.A. Johnston\address{Mathematics Department, Heriot-Watt University, 
        Edinburgh, EH14 4AS, United Kingdom}
}
       
\begin{document}

\begin{abstract}
We perform extensive Monte Carlo simulations of the 10-state Potts model
on quenched two-dimensional $\Phi^3$ gravity graphs to study the effect of
quenched connectivity disorder on the phase transition, which
is strongly first order on regular lattices. The numerical data provides strong
evidence that, due to the quenched randomness, the discontinuous first-order
phase transition of the pure model is softened to a continuous transition.

\end{abstract}

\maketitle

\section{INTRODUCTION}

Systems subject to quenched random disorder often show a completely different
behaviour than the pure case. If the pure system has a continuous phase
transition it is well known that quenched random disorder can drive the
critical behaviour into a new universality class, or the transition can even
be eliminated altogether \cite{quench}. In the case of a first-order 
phase transition in the pure system the effect of quenched random disorder
can also be very dramatic, with the possibility for a softening to a 
continuous transition.

The paradigm for testing the latter prediction is the two-dimensional (2D) 
$q$-state Potts model which exhibits on regular lattices for $q \ge 5$ a 
first-order transition whose strength increases with $q$. In Ref.~\cite{chen}
the effect of quenched {\em bond} disorder was investigated for the 
8-state model and it was found that the critical behaviour of the quenched
model could be well described by the Onsager Ising model universality class. 
In Ref.~\cite{javi95} the effect of quenched {\em connectivity} disorder
was studied by putting the $8$-state model on 2D Poissonian random lattices
showing that for this type of quenched disorder the transition stays 
first order.

A different sort of connectivity disorder appears in 2D gravity 
triangulations or their dual $\Phi^3$ graphs. In such models one is 
interested in the coupling of matter to 2D gravity, so the disorder
is annealed rather than quenched. Motivated by Wexler's mean field 
results for $q=\infty$ Potts models coupled to 2D gravity \cite{Wex}, 
simulations of the 10-state and 200-state Potts model coupled
to 2D gravity gave convincing evidence \cite{2DG} for a continuous 
transition.
 
As the only {\it quenched} connectivity disorder seriously investigated
to date, 2D Poissonian random lattices, showed no sign of softening for
first-order transitions it is interesting to enquire whether the salient
feature for the softening in the 2D gravity simulations is the annealed 
nature of the connectivity disorder or whether it is some intrinsic 
features of the graphs themselves. We discuss here results of a 
Monte Carlo (MC) study of the 10-state Potts model on quenched random 
lattices drawn from the equilibrium distribution of pure 2D gravity 
triangulations (or, more exactly, their dual $\Phi^3$ graphs).
%
                     \section{MODEL AND RESULTS}
%
We used the standard definition of the $q$-state Potts model, 
\begin{equation}
Z_{\rm Potts} = \sum_{\{\sigma_i\}} e^{\beta \sum_{\langle ij \rangle}
\delta_{\sigma_i \sigma_j}} \; ; \; 
\sigma_i = 1,\dots,q,
\label{eq:zpotts}
\end{equation}
where $\langle ij \rangle$ denotes the nearest-neighbour bonds of random
$\Phi^3$ graphs with  up to 10\,000 sites. For each lattice size we 
generated 64 independent replica using the Tutte algorithm \cite{tutte}, 
and performed long simulations of the 10-state model near the transition
point using the single-cluster update algorithm. After thermalization we
recorded measurements of the energy $E$ and the magnetization $M$ in 64 
time-series files.  The corresponding quantities per site are denoted in
the following by $e = E/N$ and $m = M/N$.

Given this raw data we employed standard reweighting techniques  
to compute, e.g., the specific heat, 
$C^{(i)}(\beta) = \beta^2 N \left( \langle e^2 \rangle - 
\langle e \rangle^2 \right)$,
for each replica labeled by the superindex $(i)$, and then performed the 
replica average, $C(\beta) \!=\! [C^{(i)}(\beta)] \!\equiv (1/64) \sum_i^{64} 
C^{(i)}(\beta)$, denoted by the square brackets. To perform the 
replica average at the level of the $C^{(i)}$ (and {\em not} at the level
of energy moments) is motivated by the general rule that quenched averages
should be performed at the level of the free energy and not the partition
function. Finally, we determined the maximum, 
$C_{\rm max} = C(\beta_{C_{\rm max}})$, for each lattice size and studied
the finite-size scaling (FSS) behaviour of $C_{\rm max}$ and 
$\beta_{C_{\rm max}}$. The error bars on the two quantities entering
the FSS analysis are estimated by jack-kniving over the 64 replicas.

The analysis of the magnetic susceptibility, $\chi(\beta) =
\beta N \left( [\langle m^2 \rangle - \langle m \rangle^2 ] \right)$ 
and the energetic Binder parameter
$V(\beta) = 1 - \left[ \langle e^4 \rangle \right]/3 
\left[ \langle e^2 \rangle \right]^2$,
proceeds exactly along the same lines, yielding $\chi_{\rm max}$ and
$\beta_{\chi_{\rm max}}$ as well as $V_{{\rm min}}$ and
$\beta_{V_{{\rm min}}}$.  
In order to be prepared for the possibility of a second-order phase
transition, the magnetic Binder parameter,
$U(\beta) = 1 -
\left[ \langle m^4 \rangle \right] /3 \left[ \langle m^2 \rangle \right]^2$,
was also measured as, in this case, its crossing 
for different lattice sizes provides an alternative
determination of the critical coupling $\beta_c$, and the 
FSS of either the maximum slopes or the slopes at 
$\beta_c$ can be used to extract the correlation length exponent.

If the transition was of first-order one would
expect for large system sizes an asymptotic FSS behaviour of the 
form $C_{\rm max} = a_C + b_C N + \dots$, and 
$\chi_{\rm max} = a_{\chi} + b_{\chi} N + \dots$. However
a linear scaling with $N$ is {\it not} consistent with our data. 
Furthermore, at a first-order phase transition one would expect that
the energetic Binder-parameter minima approach in the infinite-volume 
limit a non-trivial value related to the latent heat. Our data, however,
approaches the trivial limit of $2/3$, also strong evidence for a 
continuous transition.
\vspace{1.6in}   
\begin{figure}[h]
\vspace{1.6in}
\includegraphics{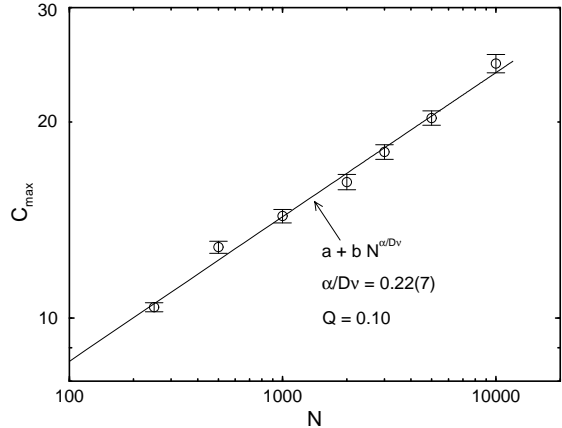}
\vspace{0.3in}
\caption{Maxima of $C$ vs $N$. The fit line is discussed
in the text.}
\label{fig1:}
\end{figure}

Being thus convinced that the transition is continuous, the next goal is
to determine the critical exponents and the corresponding universality class.
To this end we have tried to describe the scaling of the specific-heat and 
susceptibility maxima with the standard FSS ansatz
$C_{\rm max} = a_C + b_{C} N^{\alpha/D \nu}$
and $\chi_{\rm max} = a_{\chi} + b_{\chi} N^{\gamma/D \nu}$,
where $\alpha$, $\gamma$, and $\nu$ are the usual universal critical 
exponents at a continuous phase transition, $a_{C,\chi}$ and $b_{C,\chi}$ are
non-universal amplitudes, and $D$ is the intrinsic Hausdorff dimension of
the graphs.
By performing a non-linear three-parameter fit to the specific-heat maxima
we obtained $\alpha/D \nu = 0.22(7)$, with a reasonable goodness-of-fit 
parameter $Q=0.10$; see Fig.~1. 
Since the background term $a_C$ turned out to be
consistent with zero, we also tried linear two-parameter fits with 
$a_C=0$ kept fixed and the resulting exponent estimate
over all data points gave a fully
consistent value of $\alpha/D \nu = 0.222(7)$ (with $Q=0.15$) but, 
as expected, a much smaller error bar.
The susceptibility maxima grow very fast with $N$, such that also here the
constant term $a_{\chi}$ can safely be neglected. The linear fit over
all data points yielded $\gamma/D\nu = 0.732(10)$ (with $Q=0.27$), and
omitting the $N=250$ point we obtained $\gamma/D \nu = 0.719(14)$ (with
$Q=0.31$). 

The pseudo-transition points $\beta_{C_{\rm max}}$, $\beta_{\chi_{\rm max}}$,
and $\beta_{V_{{\rm min}}}$ were fitted to
the standard FSS ansatz   
$\beta_{C_{\rm max}} = \beta_c + c_C N^{-1/D\nu}$,
etc. By taking the average of the three estimates of $\beta_c$ 
we estimated the transition point to be $\beta_c = 2.2445(20)$.
This value is consistent with the estimate obtained from the crossings
of the magnetic Binder parameter $U$ for different lattice sizes.
The estimate of $1/D\nu$ from the FSS of the pseudo-transition points
is not very stable but a more precise estimate can be obtained by
analyzing the FSS of the magnetic Binder-parameter slopes in the
vicinity of $\beta_c$. Both the maximum slopes and the slopes at
$\beta_c$ are expected to scale as $d U/d\beta \propto N^{1/D\nu}$. 
The fit to the maximum slopes for $N \ge 500$
gave $1/D\nu = 0.616(29)$ (with $Q=0.29$), and for the slopes at 
$\beta_c=2.2445$ we obtained $1/D\nu = 0.614(30)$ (with $Q=0.78$).

On annealed gravity graphs it was found that the 10-state Potts model
appeared to display the exponents of the 4-state Potts model coupled
to gravity, which are listed as ``annealed $q=4$'' in Table~1 below.
The values for the exponents measured in our simulations appear to be
best fitted by the ``quenched $q=4$'' exponents derived 
in \cite{quench_theory} and also listed in Table~1.
\begin{table}[hbt]
\setlength{\tabcolsep}{0.38pc}
\newlength{\digitwidth} \settowidth{\digitwidth}{\rm 0}
\catcode`?=\active \def?{\kern\digitwidth}
\caption[{\em Nothing.}]
{Critical exponents for Potts models}
\label{tab:exponents}
\begin{tabular}{|c|c|c|c|}
\hline
Type& $\alpha /D \nu$  & $\gamma /D \nu$  & $1 /D \nu$ \\[.05in]
\hline
Annealed $q=4$ & $0$  & $1/2$ & $1/2$ \\[.05in]
\hline
Quenched $q=4$ & $0.177...$  & $0.709...$ & $0.589...$  \\[.05in]
\hline
Measured $q=10$ & $0.22(1)$ & $0.719(14)$ & $0.61(3)$ \\[.05in]
\hline
\end{tabular}
\end{table}
\noindent
On the basis of the numerical evidence in this paper
and the earlier results of \cite{2DG} it would thus seem that
the connectivity disorder of both quenched and annealed
$\Phi^3$ gravity graphs has the effect of softening the 
first-order transition of the $q=10$ Potts model to a
continuous transition. Unlike the case of bond disorder,
however, we find exponents associated with the 4-state Potts
model rather than the Ising model. It is perhaps worth remarking that 
in \cite{2DG} the 200-state Potts model on annealed gravity graphs
{\it did} apparently display Ising-like exponents,
so a similar investigation on quenched graphs might be interesting.

\section{CONCLUSIONS}
To summarize, we have obtained strong numerical evidence that 
due to connectivity disorder the phase transition in the 10-state 
Potts model on quenched random gravity graphs is softened to a 
continuous phase transition. This result is in contrast to a recent
simulation of the 8-state model on Poissonian Delaunay/Voronoi
random lattices where the transition stays first order as on regular 
lattices \cite{javi95}. It is, however, in qualitative agreement with
the quenched random bond case \cite{chen}, and like the simulations 
in \cite{2DG} of the 10-state Potts model on annealed 2D gravity graphs
we see exponents associated with the 4-state Potts model, in this case
the ``quenched'' exponents. 
\section*{ACKNOWLEGDEMENTS}
Work supported in part by EC HCM network grant ERB-CHRX-CT930343 and
NATO collaborative research grant CRG951253 (C.F.B. and D.A.J.).
W.J. thanks the DFG for a Heisenberg fellowship.
The simulations were performed on a T3D parallel computer
of Zuse Institut Berlin.
%

\end{document}